\documentclass[traditabstract]{aa}
\usepackage{natbib,graphicx}
\usepackage{amsmath,amsfonts,amssymb}
\usepackage{txfonts}
\usepackage{longtable}
\usepackage{soul,color,lscape}
\usepackage[breaklinks,colorlinks,citecolor=blue]{hyperref}

\usepackage{todonotes,ifthen}
\newcounter{mycomment}
\newcommand{\mycomment}[2][]{%
    \refstepcounter{mycomment}%
    \ifthenelse{\equal{#1}{NM}}
    {\todo[color={red!40},size=\small, inline]{%
     \textbf{{#1}\themycomment:}~#2}%
    }{\ifthenelse{\equal{#1}{JHD}}
        {\todo[color={blue!40},size=\small, inline]{%
        \textbf{{#1}\themycomment:}~#2}%
        }{\ifthenelse{\equal{#1}{EJ}}
        {\todo[color={green!40},size=\small, inline]{%
        \textbf{{#1}\themycomment:}~#2}%
        }{}}}\noindent}

\newcommand{\magneticum}{\textit{Magneticum}}
\newcommand{\Msun}{M_{\odot}}
\newcommand{\new}[1]{#1}

\usepackage[normalem]{ulem}

\begin{document}

\title{Impact of baryons in cosmic shear analyses with tomographic aperture mass statistics}

\titlerunning{Impact of baryons in cosmic shear analyses}
\authorrunning{Martinet, Castro, Harnois-D\'eraps et al.}

\author{Nicolas Martinet$^{1}$, Tiago Castro$^{2,3,4,5}$, Joachim Harnois-D\'eraps$^{6,7}$, Eric Jullo$^{1}$, Carlo Giocoli$^{8,9}$, Klaus Dolag$^{10, 11}$}

\institute{$^{1}$ Aix-Marseille Univ, CNRS, CNES, LAM, Marseille, France\\
$^{2}$ Dipartimento di Fisica, Sezione di Astronomia, Universit\`a di Trieste, Via Tiepolo 11, I-34143 Trieste, Italy\\
$^{3}$ INAF -- Osservatorio Astronomico di Trieste, via Tiepolo 11, I-34131 Trieste, Italy\\
$^{4}$ IFPU -- Institute for Fundamental Physics of the Universe, via Beirut 2, 34151, Trieste, Italy\\
$^{5}$ INFN -- Sezione di Trieste, I-34100 Trieste, Italy\\
$^{6}$Astrophysics Research Institute, Liverpool John Moores University, 146 Brownlow Hill, Liverpool, L3 5RF, UK\\
$^{7}$ Institute for Astronomy, University of Edinburgh, Royal Observatory, Blackford Hill, Edinburgh EH9 3HJ, UK\\
$^{8}$  INAF - Osservatorio di Astrofisica e Scienza dello Spazio di Bologna, via Gobetti 93/3, I-40129 Bologna, Italy\\
$^{9}$  INFN - Sezione di Bologna, viale Berti Pichat 6/2, I-40127 Bologna, Italy\\
$^{10}$ University Observatory Munich, Scheinerstr. 1, 81679 Munchen, Germany\\
$^{11}$ Max-Planck Institut fur Astrophysik, Karl-Schwarzschild Str. 1, D-85741 Garching, Germany\\
  \email{nicolas.martinet@lam.fr}}

\setcounter{page}{1}

\abstract{NonGaussian cosmic shear statistics based on weak-lensing aperture mass ($M_{\rm ap}$) maps can outperform the classical shear two-point correlation function ($\gamma$-2PCF) in terms of cosmological constraining power. However, reaching the full potential of these new estimators requires accurate modeling of the physics of baryons as the extra nonGaussian information mostly resides at small scales. We present one such modeling based on the \textit{Magneticum} hydrodynamical simulation for the KiDS-450 and DES-Y1 surveys and a \textit{Euclid}-like survey. We compute the bias due to baryons on the lensing PDF and the distribution of peaks and voids in $M_{\rm ap}$ maps and propagate it to the cosmological forecasts on the structure growth parameter $S_8$, the matter density parameter $\Omega_{\rm m}$, and the dark energy equation of state $w_0$ using the SLICS and cosmo-SLICS sets of dark-matter-only simulations. We report a negative bias of a few percent on $S_8$ and $\Omega_{\rm m}$ and also measure a positive bias of the same level on $w_0$ when including a tomographic decomposition. These biases reach $\sim 5$\% when combining $M_{\rm ap}$ statistics with the $\gamma$-2PCF as these estimators show similar dependency on the AGN feedback. We verify that these biases constitute a less than $1\sigma$ shift on the probed cosmological parameters for current cosmic shear surveys. However, baryons need to be accounted for at the percentage level for future Stage IV surveys and we propose to include the uncertainty on the AGN feedback amplitude by marginalizing over this parameter using multiple simulations such as those presented in this paper. Finally, we explore the possibility of mitigating the impact of baryons by filtering the $M_{\rm ap}$ map but find that this process would require to suppress the small-scale information to a point where the constraints would no longer be competitive.}


\maketitle

\section{Introduction}
\label{sec:intro}

Weak lensing cosmic shear is one of the most powerful cosmological probes of the late-time Universe. So far, most analyses have focused on studying the correlation between the shape distortions of pairs of galaxies as a function of their separation: the so-called shear two-point correlation function \citep[$\gamma$-2PCF; e.g., ][]{Troxel+18,Hikage+19,Asgari+20}. However, it is becoming clear that other estimators, and in particular those based on weak-lensing mass maps, outperform the standard $\gamma$-2PCF \citep[e.g.,][]{DH10,Ajani+20,Zurcher+20,Coulton+20,Martinet+20}. Indeed, mass maps are highly sensitive to the nonGaussian part of the matter distribution that arises from the nonlinear growth of structures, which contains information that is overlooked by two-point estimators. Consequently, the combination of both probes yields even tighter constraints, as seen in recent applications to observational data \citep[e.g.,][]{Martinet+18,Harnois-Deraps+20}. With future Stage IV surveys, this combination is also expected to improve not only our measurement of the growth of structure parameter $S_8$ by a factor of two, but also that of the dark energy equation of state $w_0$ \citep{Martinet+20} and of the sum of neutrino masses $\Sigma m_\nu$ \citep{Li+19}, by factors of three and two, respectively.

Nevertheless, these nonGaussian estimators are difficult to predict theoretically because of limits in our understanding of the nonlinear growth of structures \citep[see e.g.,][for some attempts]{Fan+10,Lin+15,Shan+18,Giocoli+18b,Barthelemy+20b} and are instead modeled with $N$-body simulations. This can significantly increase the computational cost of such analyses, but resorting to $N$-body simulations is also necessary to accurately model the $\gamma$-2PCF at scales affected by nonlinearities \citep[e.g.,][]{EuclidIX}. Moreover, many public simulation suites can be exploited for this purpose, including the Scinet LIght-Cone Simulations \citep[][SLICS]{Harnois-Deraps+15}, the cosmo-SLICS \citep{cosmoSLICS}, or the MassiveNuS \citep{Liu+18}.

As mass map estimators focus on small scales of typically around a few arcminutes, they can be severely affected by baryonic feedback, which can bias the inferred cosmological constraints. To avoid this, an effective approach is to quantify the impact of baryons on the estimator with hydrodynamical simulations, and to apply a correction factor to the model extracted from dark matter (DM) only simulations. For mass maps, this approach was pioneered in \citet{Yang+13} and \citet{Osato+15}. However, active galactic nuclei (AGN) feedback was not included in the first analysis and only with a low amplitude of the feedback in the second. As a result, these two studies found a mild impact of the baryonic physics on the distribution of peaks in mass maps and therefore underestimated the bias on cosmological parameters \citep[e.g.,][]{Weiss+19,Coulton+20}.

State-of-the-art hydrodynamical simulations with box lengths of hundreds of megaparsecs and including realistic AGN feedback later enabled refinement of the measure of the bias on mass map estimators due to baryons. \citet{Fong+19} measured a $\sim10$\% reduction in the number of high signal-to-noise-ratio (S/N) $M_{\rm ap}$ peaks in the BAHAMAS simulations \citep{McCarthy+17}, with a particular look at possible degeneracies between the effect of baryons and that of massive neutrinos. \citet{Osato+20} measured biases of the same order of magnitude on the number of peaks, the number of minima, and the lensing probability distribution function (PDF) in the TNG 300 simulation \citep{Pillepich+18}, also highlighting a less pronounced bias at higher redshift.

An interesting alternative to hydrodynamical simulations is the baryonification method described in \citet{Schneider+15a} where particle positions are shifted in DM-only simulations to mimic the impact of baryons. This method now accurately reproduces AGN feedback and star formation compared to hydrodynamical simulations \citep{Arico+20} and could offer an efficient way of decreasing the computational resources needed to include baryonic effects in cosmological models. This technique is however not tested in the present article as it requires the particle positions, which are generally not stored for a posteriori applications. \citet{Weiss+19} applied this baryonification method to model the impact of baryons on peak statistics and also noted that the latter could be mitigated by smoothing the small scales by applying a Gaussian filter to the mass map. This smoothing is nevertheless likely to also reduce the statistical power of the mass map estimators, a hypothesis that can only be verified by propagating this bias to the cosmological parameter inference.

Recently, \citet{Coulton+20} performed this propagation and measured the impact of baryons directly on the forecasts of the matter density $\Omega_{\rm m}$, the amplitude of fluctuations $A_{\rm S}$, and the sum of the neutrino masses for peaks and minima using the BAHAMAS simulations. In a nontomographic approach, these latter authors found larger biases for peaks than for minima in their LSST-like mock data, concluding that the latter is potentially more robust against baryons.

Building on these previous analyses, and exploiting the cosmological analysis pipeline introduced in \citet{Martinet+20}, we measure for the first time the effect of baryons on the dark matter and dark energy cosmological parameters in a tomographic Stage-IV lensing survey setup. We focus on the particular case of aperture-mass \citep[$M_{\rm ap}$;][]{Schneider96} maps, which are particularly well suited for cosmological analyses. We model the effect of baryons with the \magneticum\footnote{\url{www.magneticum.org}} hydrodynamical simulation suite \citep[e.g.,][]{Castro+21}, which includes all key ingredients about the physics of baryons, such as AGN feedback \citep{Springel+05b,Fabjan+10,Hirschmann+14} and star formation \citep{Springel+03}. Both effects redistribute the matter in and around DM haloes, but the exact amplitude of the feedback varies between simulations \citep[see][for a recent review on feedback in cosmological hydrodynamical simulations]{Chisari+19}. Extending the simulation suites of \citet{Martinet+20} based on the SLICS and the cosmo-SLICS, we construct \textit{Euclid}-like mocks from the \magneticum\ hydrodynamical simulation. We measure the impact of baryons on the $\gamma$-2PCF, on the lensing peaks, minima, and on the lensing PDF, in the form of a multiplicative baryon bias correction factor. We then propagate this correction into the cosmological inference pipeline described in \citet{Martinet+20}, and investigate the impact on the parameter forecasts. Finally, we explore various mitigation schemes to decrease the baryon-dominated small-scale contribution to the $M_{\rm ap}$ computation.

We introduce the \magneticum\ simulation and compare it to other state-of-the-art hydrodynamical simulations in Sect.~\ref{sec:sim}. We present the methodology that we employ to measure the bias due to baryons in Sect.~\ref{sec:met}. We then measure their impact on different data vectors (DV) in Sect.~\ref{sec:imp;subsec:dv} and propagate the effect to the cosmological forecasts in Sect.~\ref{sec:imp;subsec:cp}. We test different mitigation setups in Sect.~\ref{sec:cor} and conclude in Sect.~\ref{sec:ccl}. Finally, we adapt our mocks to the KiDS and DES surveys in Appendix~\ref{sec:app} and measure the impact of baryons on the cosmological constraints by \citet{Martinet+18,Harnois-Deraps+20} in these two surveys.

\section{Modeling baryonic effects}
\label{sec:sim}

\subsection{The \magneticum\ hydrodynamical simulation}

\begin{figure}
    \centering
    \includegraphics[width=0.48\textwidth]{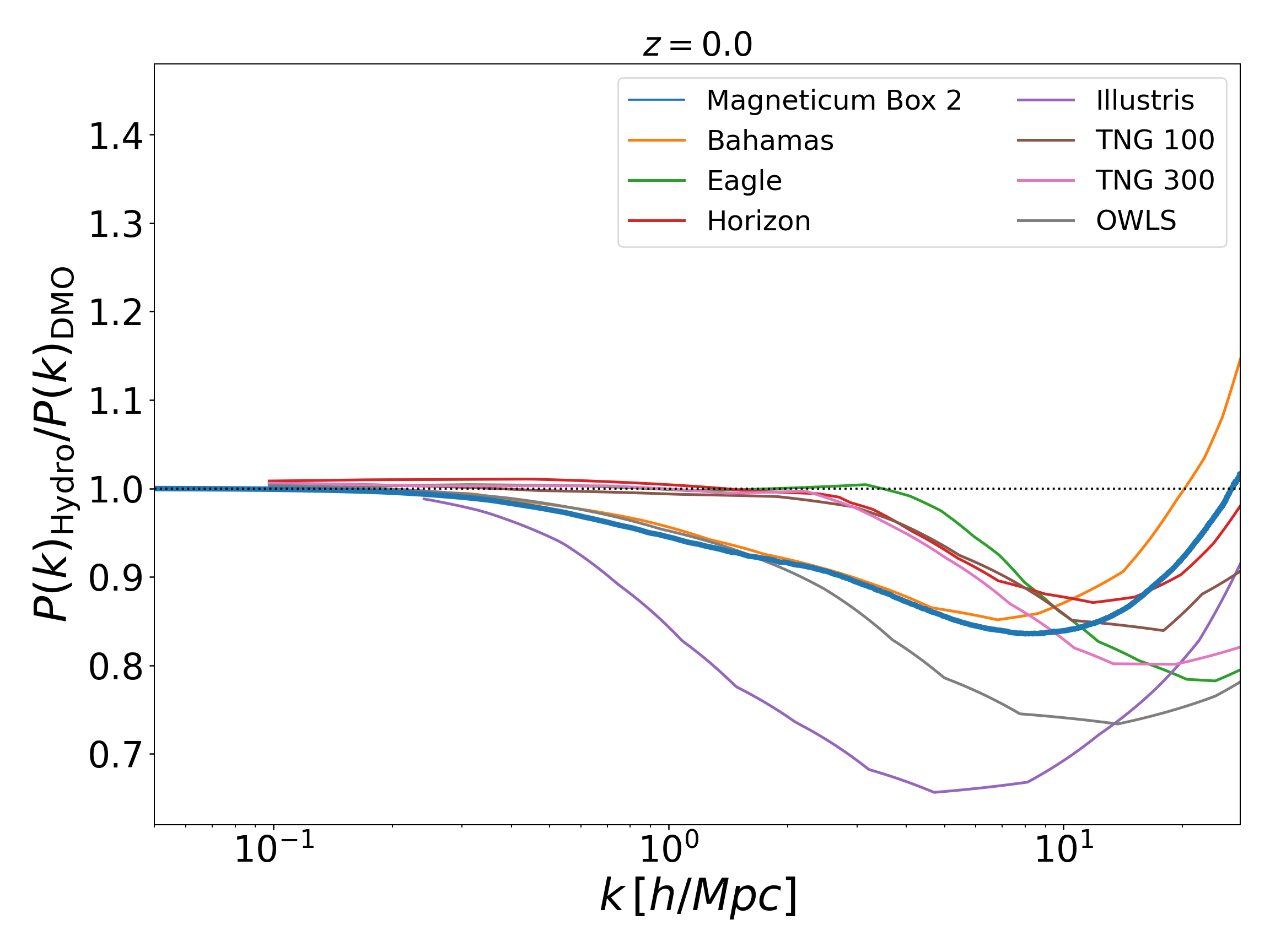}
    \caption{Ratio of power spectra at $z=0$ between various hydrodynamical simulations and their corresponding DM-only run. The \magneticum\ used in this analysis is representative of an average behavior with a loss of power of about $15$\% at $k\sim10 ~ h\,{\rm Mpc}^{-1}$ due to AGN feedback.}
    \label{fig:Pk}
\end{figure}

\begin{table*}
\begin{minipage}{\textwidth}
\begin{center}
{
\caption{Subset of the \magneticum\ simulation suite used in this work. From left to right: box size,  gravitational softening and the particle masses for the different components (DM, gas, and stars), the number of lens planes built, redshift range of the past-light cone, its field of view, and the map angular resolution. A ``$-$'' indicates that the parameter value of Box 2b/hr is identical to that of Box 2/hr.}
\setlength{\tabcolsep}{4.0pt}
\begin{tabular}{lcccccccccccccc}
    \hline\hline
    Box  & $L_\textrm{box}$ & \multicolumn{3}{c}{$\epsilon_{\rm{soften.}}$(kpc$\,h^{-1}$)} & $N_\textrm{particles}$ & $m_\textrm{DM}$ & $m_\textrm{gas}$ & $m_\textrm{star}$ & $N_{\rm planes}$ &  $z_{\rm min.}$ & $z_{\rm max.}$ & FoV & Pixel Size \\\cline{3-5}
    name & {\rm (Mpc$\,h^{-1}$)}& DM & Gas & Stars & & $(\Msun\,h^{-1})$   & $(\Msun\,h^{-1})$    & $(\Msun\,h^{-1})$  &  & & & (deg.) & (arcsec.)\\
    \hline
  2/hr & $352$& $\;\,3.75$ & $3.75$ & $2.0$ & $2 \times1584^3$& $6.9\times10^8$ & $1.4\times10^8$ & $2.3\times10^7$ & $4$ & $0.0$ & $0.248$  & $10.0$ & $3.6$\\
  2b/hr & $640$& $\;\,-$ & $-$ & $-$ & $2 \times2880^3$& $-$ & $-$ & $-$ & $13$ & $0.248$ & $3.44$ & $-$ & $-$\\
 \hline
\end{tabular}
}
\label{tab:sims}
\end{center}
\end{minipage}
\end{table*}

The \magneticum\ suite  \citep{2013MNRAS.428.1395B, Saro:2013fsr, 2015MNRAS.448.1504S, 2016MNRAS.458.1013S, Dolag:2014bca, Dolag:2015dta, 2015ApJ...812...29T, Remus:2017dns, Castro+21} is a compilation of $N$-body and hydrodynamical simulations describing the cosmic evolution of the Universe. In total, the suite follows up to $2\times10^{11}$ particles divided in DM, gas, stars, and black holes. The simulations were performed with the TreePM+SPH code P-Gadget3 --- a higher performance and more efficient version of the publicly available Gadget-2 code~\citep{Springel05} developed concomitantly to its successor Gadget-4~\citep{Springel:2020plp}. Notably, the smoothed-particle-hydrodynamics (SPH) solver implements the improved model of~\citet{beck:2015qva}. The particle dynamics is coupled to different astrophysical effects such as radiative cooling, heating by a uniform evolving UV background, star formation~\citep{Springel+03}, stellar evolution, and chemical enrichment processes~\citep{tornatore:2007ds}. Cooling is implemented following the metallicity-dependent formulation presented in~\cite{wiersma:2008cs} using cooling tables produced by the publicly available CLOUDY photo-ionization code~\citep{ferland:1998id}. Lastly, AGN feedback and black hole growth are modeled as described in~\citet{michaela:2013sia}.

The sub-grid physics model of our simulation reproduces a long list of observations, from galaxy properties~\citep{2015ApJ...812...29T,Remus:2016elq} and AGN population~\citep{Hirschmann+14,2016MNRAS.458.1013S}, to the intergalactic and intercluster medium~\citep{Dolag:2015dta,bocquet:2015pva,Gupta:2016yso}. Of particular importance for this work is the robustness of the AGN feedback of our model, which has a clear footprint on the matter power-spectrum, suppressing its amplitude with respect to the DM-only case on $k \! \sim \! 10\,h\,{\rm Mpc}^{-1}$ by $\sim \! \! 15 \%$. The specific suppression range and amplitude depend strongly on the AGN model. In Fig.~\ref{fig:Pk} we compare the ratio of the matter power-spectrum computed from Box $2$ (see Table~\ref{tab:sims} for details on the simulations) Hydro and DM-only with other simulations presented in \citet{Chisari+19}: BAHAMAS \citep{McCarthy+17}, EAGLE \citep{Schaye+15}, Horizon \citep{Dubois+14}, Illustris \citep{Vogelsberger+14}, IllustrisTNG \citep{Pillepich+18}, and OWLS \citep{Schaye+10}. As can be observed, \magneticum\ provides an AGN feedback suppression consistent with other simulations, in particular with BAHAMAS and to a lesser extent with TNG-300, which were used in other recent mass map analyses. We note that the AGN feedback also controls the gas fraction of halos (and the conversion efficiency into stars) such that the simulations presented in Fig.~\ref{fig:Pk} also differ in these quantities although the matter power spectrum is a good indicator with which to assess the current theoretical uncertainties in modeling baryonic physics.
%

\subsection{ Past light-cone reconstruction}

The \magneticum\ past light-cone reconstructions are performed using the Simulation LIght conE buildeR code --- \textsc{SLICER}\footnote{\href{https://github.com/TiagoBsCastro/SLICER}{https://github.com/TiagoBsCastro/SLICER}} --- closely following the pipeline presented in~\citet{Castro:2017tbn} and forked from a MapSim branch \citep{Giocoli+15,Giocoli+18a,Hilbert+20}. Briefly, light-cones are built in post-processing, assigning particles to predetermined $2$D mass maps according to the triangular-shaped cloud mass assignment scheme. The geometry of the past light-cones is a square-based pyramid (in angular coordinates) where the observer is located at the $z=0$ vertex, and which extends to $z_{\rm max}$.  The opening angle is chosen to be $10$ degrees and the angular resolution of the light-cone mass planes is $3.6$ arcsec. \textsc{SLICER} allows mass maps thicknesses that are a rational fraction of the box size, which we have chosen to be half of the box size. Every two mass maps, particles are randomly shifted and reflected with respect to one of the box axes (accounting  for  periodic  boundary  conditions) in order to avoid the repetition of the same cosmic structure along the line of sight. 

Subsequently, mass maps are converted into surface density maps $\Sigma(x,y)$ as
\begin{equation}
    \Sigma(x,y) = \dfrac{\sum_{j=1}^{n} m_{j}}{L_p^2}\,,
\end{equation}
where $n$ indicates the number of particles, $m_j$ the interpolated contribution of the $j^{\rm th}$ particle to the pixel at position $x,y,$ and $L_p$ is the physical size of the pixel in units of Mpc$\,h^{-1}$. Given a source plane, convergence maps $\kappa(x, y)$ are created by weighting these maps by the corresponding critical surface mass density,
\begin{equation}
    \Sigma_{\rm crit} \equiv \dfrac{c^2}{4 \pi G} \dfrac{D_l}{D_s D_{ls}},
\end{equation}
and integrating over the past light-cone. Here, $c$ indicates the speed of light, $G$ the Newton's constant, and $D_l$, $D_s$, and $D_{ls}$ are the angular diameter distances between observer--lens, observer--source, and lens--source, respectively. The shear components ($\gamma_1$, $\gamma_2$) are obtained from the standard inversion technique of \citet{KS93}.

As Box $2$b Hydro has not been run down to $z=0$, our past light-cones are built from grafting Box $2$ in the range $z=[0,0.248]$ with Box $2$b for $z=[0.248,3.44]$. We used the DM-only runs to validate our approach, and verified that for the source redshift distribution of interest in this paper, grafting only marginally affects the lensing statistics for $z_s\gtrsim1.0$ with respect to the same statistics inferred from a contiguous light-cone\footnote{We note that a similar grafting strategy has been used for the high-precision ``Clone'' data used by the CFHTLenS team \citep{CFHTLenSClone}}. In total, we produced $20$ \textit{pseudo-independent} light-cones --- $10$ Hydro and $10$ DM-only --- with properties summarized in Table~\ref{tab:sims}. 

\section{Methodology}
\label{sec:met}

In this section, we first give a brief description of our cosmological forecasting  pipeline in Sect.~\ref{sec:met;subsec:cos}, and then detail the calculation of baryonic bias in Sect.~\ref{sec:met;subsec:bias}.

\subsection{From shear to cosmology}
\label{sec:met;subsec:cos}

Cosmological forecasts are computed with the same pipeline as in \citet{Martinet+20}. While we recapitulate the salient points of this analysis here, we refer the reader to this publication for more details.

\begin{itemize}

    \item[$\bullet$] {\it $w$CDM Simulations:} the cosmology dependence of $M_{\rm ap}$ statistics is emulated with radial basis functions based on measurements from Stage-IV mock data covering $25$ cosmologies in $S_8 - \Omega_{\rm m} - w_0 - h$, organised in a latin hypercube. These were built from  the cosmo-SLICS $N$-body simulations \citep{cosmoSLICS} and contain $50$ mocks per cosmology, covering ten light-cones and five shape noise realizations to further increase our precision on the model. The covariance is measured from a separate suite of \new{$N_{\rm s}=928$} fully independent $\Lambda$CDM mocks  extracted from the SLICS $N$-body simulations \citep{SLICS18} with the same survey properties. These Stage-IV mocks match the expected $30$ arcmin$^{-2}$ galaxy density of \textit{Euclid} with a realistic redshift distribution in the range $0<z<3$ \citep{Laureijs+11}, and an opening angle of $100$ deg$^2$ each. While galaxy positions and intrinsic ellipticities are chosen randomly in the covariance mocks, they are fixed across the different cosmologies in cosmo-SLICS mocks so as to reduce the impact of shape noise on the model. \\

    \item[$\bullet$] {\it Measurements}: From each mock, we compute a $1024\times1024$ pixel $M_{\rm ap}$ map \citep{Schneider96} as a convolution between the tangential ellipticity $\epsilon_{\rm t}$ around a pixel $\vec{\theta}_0$, and the \citet{Schirmer+07} compensated $Q$ filter adapted to the detection of matter halos:
    \begin{equation}
    \label{eq:map}
    M_{\rm ap}(\vec{\theta}_0)=\frac{1}{n_{\rm gal}}\sum_i Q(|\vec{\theta}_i-\vec{\theta}_0|)\,\epsilon_{{\rm t}}(\vec{\theta}_i, \vec{\theta}_0), 
    \end{equation}
    with
    \begin{equation}
    \label{eq:Q}
    \begin{aligned}
    Q(\theta) =  & \left[1 + \exp \left(150\frac{\theta_{\rm in} - \theta}{\theta_{\rm ap}}\right) + \exp \left(-47 +50 \frac{\theta}{\theta_{\rm ap}}\right)\right]^{-1} \\
             & \times \left(\frac{\theta}{x_{\rm c}\theta_{\rm ap}}\right)^{-1} \tanh \left(\frac{\theta}{x_{\rm c}\theta_{\rm ap}}\right).
    \end{aligned}
    \end{equation}
    This filter is a simpler form of the NFW \citep{NFW97} profile with an additional exponential attenuation in the center and at the edge of the aperture. $x_{\rm c}$ controls the tilt between the core and the edge of the profile and we introduce an optional inner radius parameter $\theta_{\rm in}$ to govern the exponential cut-off in the inner part. The sum in Eq. (\ref{eq:map}) is carried out over galaxies at position $\vec{\theta}_i$ within an aperture of radius $\theta_{\rm ap}=10'$ by default (corresponding to an effective smoothing scale of $\theta_{\rm ap} x_{\rm c} = 1.5'$) and centered on $\vec{\theta}_0$, and is normalized by the galaxy density $n_{\rm gal}$ within the aperture. We estimate the noise due to intrinsic ellipticities in the mocks to define several $M_{\rm ap}$ map statistics based on the S/N distribution of pixels of these maps: peaks and voids (pixels with values greater or smaller than their $8$ neighbors), as well as the full distribution of pixels (1D $M_{\rm ap}$, often referred to as the lensing PDF). Following \citet{Martinet+20}, these distributions are organised in eight bins between $-2.5<{\rm S/N}<5.5$ for peaks, eight bins between $-5<{\rm S/N}<3$ for voids, and in nine bins between $-4<{\rm S/N}<5$ for the 1D $M_{\rm ap}$. For each mock, we also compute the $\gamma$-2PCF $\xi_\pm(\vartheta)$ using the \textit{Athena} software \citep{Kilbinger+14}, with the estimators  defined as \citep{Schneider+02}:
    $\xi_\pm(\vartheta) = \left[\sum_{ij} \epsilon_{\rm t}^{i}\epsilon_{\rm t}^{j} \pm \epsilon_{\times}^{i}\epsilon_{\times}^{j}\right] / N_{\rm pairs}(\vartheta)$, where the sum is over galaxy pairs $ij$ separated by an angle $\vartheta$ and potentially in different tomographic bins. The results are binned in eight angular bins logarithmically spaced between $0.1'$ and $60.5'$ for $\xi_+$ and between $0.5'$ and $300'$ for $\xi_-$. The quantities $\epsilon_{\rm t, \times}$ are the tangential and ``cross'' components of the ellipticities.\\

    \item[$\bullet$] {\it Tomography}: We separate the mock data into five redshift bins: $0<z_1<0.47$, $0.47<z_2<0.72$, $0.72<z_3<0.96$, $0.96<z_4<1.33$, and $1.33<z_5<3$. We reconstruct  $M_{\rm ap}$ maps from the galaxy samples in individual redshift slices (auto-$M_{\rm ap}$) and from the combination of multiple redshift bins from two and up to five \citep[the cross-$M_{\rm ap}$ terms introduced in][]{Martinet+20}. The latter showed that including the cross-maps yields a significant improvement in the forecast precision. For the $\gamma$-2PCF, we include both the auto- and the cross-tomographic correlations.\\

    \item[$\bullet$] {\it Likelihood}: Finally, we compute the likelihood \new{of the data  given the cosmology $p(\vec{x}|\vec{\pi})$} on a four-dimensional grid, taking the ``fiducial'' cosmo-SLICS simulations as our \new{observation DV $\vec{x}$}. We use a Student-$t$ likelihood \citep{Sellentin+16},
    \begin{equation}
    \label{eq:like}
    \new{p(\vec{x}|\vec{\pi}) \propto \left[ 1+\frac{\chi^2\left(\vec{x},\vec{\pi}\right)}{N_{\rm s}-1}\right]^{-N_{\rm s}/2},}
    \end{equation}
    \begin{equation}
    \label{eq:chi2}
    \new{\chi^2(\vec{x},\vec{\pi}) = (\vec{x}-\vec{x}_{\rm m}(\vec{\pi}))^{\rm T} \; \Sigma^{-1}(\vec{\pi}_0)  \; (\vec{x}-\vec{x}_{\rm m}(\vec{\pi})),}
    \end{equation}
    which generalizes the multi-variate Gaussian, and we account for the noise through the covariance matrix \new{$\Sigma$ computed at a fixed cosmology $\vec{\pi}_0$. In the above equation, $\vec{x}_{\rm m}(\vec{\pi})$ refers to the DV modeled from the cosmo-SLICS at cosmology $\vec{\pi}$ and both $\vec{x}$ and $\vec{x}_{\rm m}(\vec{\pi})$ correspond to the mean DV over the different noise realizations. The posterior likelihood is obtained from Bayes' theorem with the cosmo-SLICS parameter space as a fixed prior ($p(\vec{\pi})=1$ within that space and $0$ outside)}. We also improve the accuracy of the emulator used in \citet{Martinet+20}, initially limited by the number of points in the interpolation grid, by performing a second interpolation in the parameter space restricted to the hyper-volume where the likelihood is nonzero.
    We measure the $1\sigma$ uncertainty on each cosmological parameter by finding the range of values enclosing $68$\% of the \new{posterior} likelihood previously marginalized over all other parameters. As the posterior distribution on $h$ extends up to the prior limit imposed by the simulation space, we only discuss the forecasts for the other three probed parameters\footnote{Our  priors on the four cosmological parameters are given by the range of the cosmo-SLICS simulations: $\Omega_{\rm m} \in [0.1,  0.55]$, $S_8 \in [0.6, 0.9]$, $w_0 \in [-2.0,  -0.5]$ and $h\in [0.6, 0.82]$.}: $S_8$, $w_0$, and $\Omega_{\rm m}$. All predictions are computed for a $100$ deg$^2$ area to ensure that the few percent uncertainties in the model (primarily due to the accuracy of the $N$-body simulations) are lower than the precision of the forecasts.

\end{itemize}

\subsection{Measuring biases}
\label{sec:met;subsec:bias}

In this article, we use the pipeline described above to study the impact of the baryonic bias on a Stage-IV cosmic shear data analysis. However, we note  that we could instead propagate  biases due to shear measurement uncertainty, mean photometric redshift inaccuracy, galaxy intrinsic alignments, or source-lens coupling \citep[see e.g.,][]{Kacprzak+16,Harnois-Deraps+20}, but we leave this to future work. The strategy is to bias the observation DV and compare the positions of maximum likelihood to the no-bias case.

We use the DM-only and equivalent hydrodynamical runs of the \magneticum\ simulations to measure the cosmological bias due to baryonic effects. We create galaxy mocks that reproduce the same Stage-IV survey properties as the cosmo-SLICS. In particular, galaxy redshifts, positions, and intrinsic ellipticities are identical to those of the model. We generate $50$ realizations of the hydro and DM light-cones: ten different lines of sight to lower the sample variance, each populated with five different realizations of intrinsic ellipticities to converge on an average shape noise contribution. We compute the $M_{\rm ap}$ map in every mock, extract the DV, and measure the ratio between the average over the $50$ DVs measured in the hydrodynamical and in the DM-only mocks.

This multiplicative correction factor is computed for each S/N bin and serves to infuse our \magneticum\  baryon model in our (DM-only) observation data. Infusing baryons into the model avoids being affected by small residual differences between the \magneticum\ DM-only and the cosmo-SLICS DM-only simulations, such as the finite box effect \citep[e.g.,][]{Harnois-Deraps+15b,EuclidIX}, the nonlinear physics modeled by the different Poisson solvers, or the chosen distances between lensing planes \citep[e.g.,][]{Takahashi+17,Zorilla+19}. In other words, it ensures that the differences in the likelihood maxima are only due to baryons. We neglect here the possible dependence of baryons on cosmology, a hypothesis well supported by the recent findings of \citet{vanDalen2020}.

\section{Impact of baryons}
\label{sec:imp}

We examine in this section the impact of baryons on the different DVs (Sect.~\ref{sec:imp;subsec:dv}) and on the cosmology inferred by a Stage-IV survey (Sect.~\ref{sec:imp;subsec:cp}).

\subsection{Impact on computed statistics}
\label{sec:imp;subsec:dv}

\begin{figure}
    \centering
    \includegraphics[width=0.5\textwidth]{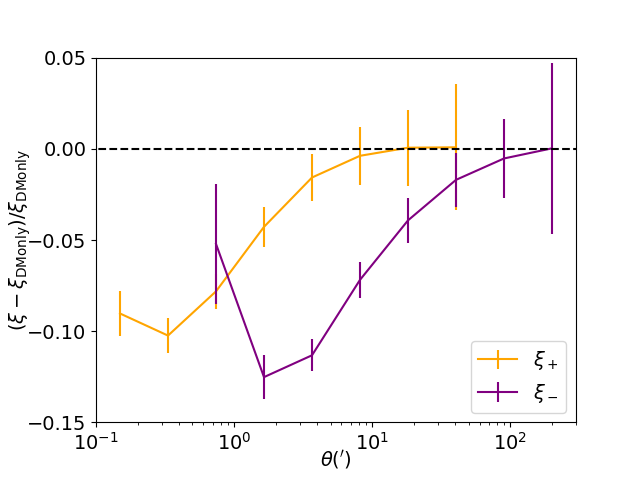}
    \caption{Relative change in the $\gamma$-2PCF due to baryons. The orange and purple curves represent the $\xi_+$ and $\xi_-$ estimators, respectively. The error bars correspond to the diagonal elements of the SLICS covariance matrix rescaled to $15\,000$~deg$^2$.}
    \label{fig:DV2pcf}
\end{figure}

Figure~\ref{fig:DV2pcf} shows the impact of baryons on the $\gamma$-2PCF, presented as the fractional difference between $\xi_\pm$ measured in the hydrodynamical and in the equivalent DM-only runs. This figure shows the baryonic bias in the absence of tomography for visual purposes; however similar curves are observed in the tomographic case. We measure a decrease in the amplitude of $\xi_\pm$ of up to $10-15$\% at scales below a few arcminutes, which is fully consistent with the suppression in the matter power spectrum seen in Fig.~\ref{fig:Pk}. 

\begin{figure}
    \centering
    \includegraphics[width=0.5\textwidth]{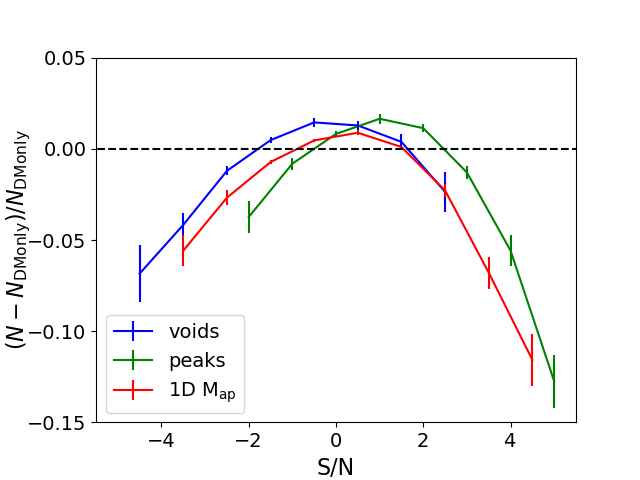}
    \caption{Relative change in the $M_{\rm ap}$ estimators due to baryons. The blue, green, and red curves represent void counts, peak counts, and the lensing PDF, respectively. The error bars correspond to the diagonal elements of the SLICS covariance matrix rescaled to $15\,000$~deg$^2$.}
    \label{fig:DV}
\end{figure}

In Fig.~\ref{fig:DV} we show the equivalent measurement on the $M_{\rm ap}$ statistics: voids, peaks, and 1D $M_{\rm ap}$. The impact of baryons on these DVs is more complex than in the case of the $\gamma$-2PCF given the different physical origins of each part of the $M_{\rm ap}$ distributions. The most plausible scenario is described in \citet{Osato+20}: Overdense regions are diluted due to AGN feedback, leading to high-S/N structures with smaller amplitudes. This expelled material can be deposited in low-density regions, which likely explains the reduction in pixels with highly negative S/N. We note that the cause of the latter effect is not yet fully understood and could involve other baryonic processes such as interaction of internal and accretion shocks \citep[e.g.,][]{Zhang+20}. The increase of the distribution at S/N close to zero accounts for the density smoothing due to this redistribution of matter which leads to a higher number of small S/N structures. This reasoning is deducted from the lensing PDF behavior but holds for peaks and voids. We do not probe the impact of radiative cooling seen at very high S/N in \citet{Osato+20} as we focus here on a lower, more conservative S/N range. This would also necessitate a  finer pixel scale than our fiducial $0.59'$. Indeed these effects appear to be significant only at scales lower than $0.5'$ in the $M_{\rm ap}$ map according to the location of the $\xi_+$ upturn in Fig.~\ref{fig:DV2pcf} and consistent with the propagation of the radiative cooling scale of $\sim 15~h\,{\rm Mpc}^{-1}$ into the lensing PDF using Eq. (11) of \citet{Castro:2017tbn}.

Quantitatively, we find a decrease of between $5$\% and $13$\% in the extremal values of our DVs depending on the considered DV and S/N bin. In particular, for peaks of $S/N=4$ we measure a depletion of $\sim 5$\% due to baryons, in perfect agreement with measurements from \citet{Fong+19} and \citet{Coulton+20} on the BAHAMAS simulations, which present a similar amplitude of the baryonic feedback (see Fig.~\ref{fig:Pk}). This is also of the same order of magnitude as the results from the TNG simulations \citep{Osato+20} and the baryonification method \citep{Weiss+19}.

When applying tomography, we also note a decrease in the impact of baryons in each slice compared to the combined case. Using thin source slices, \citet{Osato+20} also found that the baryon bias is lower at higher redshift. However, the lower effect in our case is more likely due to an increase in the noise due to lower galaxy densities as noted by \citet{Harnois-Deraps+20} in their tomographic peak count analysis of DES data. This is supported by the fact that we find similar baryonic effects in all five redshift slices, which we designed to have identical galaxy densities. 

In Figs.~\ref{fig:DV2pcf} and~\ref{fig:DV} the error bars are computed for the expected $15\,000$ deg$^2$ of the \textit{Euclid} survey by area-rescaling the SLICS covariance matrix as Cov$\;\rightarrow\;$Cov\;($100/15\,000$). These figures show that the impact of baryons on all tested estimators will be significant with respect to the statistical precision of Stage IV cosmic shear surveys. For the present $100$ deg$^2$ analysis, the changes are below the statistical noise, whence the necessity to fix the noise in the cosmo-SLICS model and the \magneticum\ mocks.

\subsection{Propagation to cosmological constraints}
\label{sec:imp;subsec:cp}

We propagate the impact of baryons on the cosmological parameter forecasts following the method described in Sect.~\ref{sec:met}. We focus on the peak statistics as it is the most widely used mass map estimator in the literature, but we also present results for voids and the lensing PDF. All values reported in this section are summarized in Table~\ref{tab:res}. 

\begin{figure}
    \centering
    \includegraphics[width=0.5\textwidth]{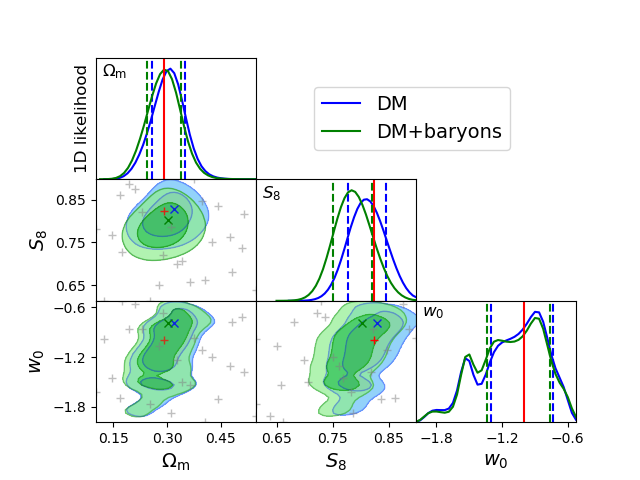}
    \caption{Forecast of cosmological parameters from peak counts in a $100$~deg$^2$ survey at \textit{Euclid} depth without tomography. Marginalized 2D ($1$ and $2\sigma$ contours) and 1D (full likelihood) constraints are displayed in blue for the DM-only case and in green when including baryon physics. Dashed lines correspond to the $1\sigma$ constraints in the 1D marginalized likelihood. The blue and green crosses indicate the best estimate in the 2D constraints\new{, while the red crosses and lines indicate the input cosmological parameter values.} Gray crosses correspond to parameters of the $25$ cosmo-SLICS simulations that are used to estimate the cosmology dependence of the number of peaks.}
    \label{fig:forecasts_8_1}
\end{figure}

\begin{figure}
    \centering
    \includegraphics[width=0.5\textwidth]{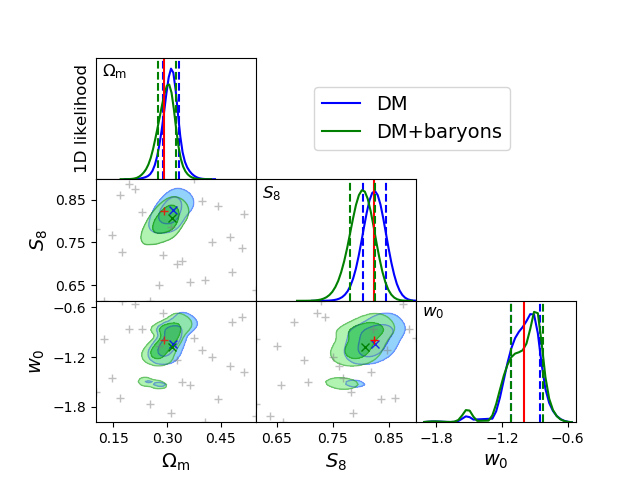}
    \caption{Same as Fig.~\ref{fig:forecasts_8_1} but for a tomographic analysis with five redshift slices and including auto- and cross-$M_{\rm ap}$ terms between redshift slices.}
    \label{fig:forecasts_8_25}
\end{figure}

\begin{figure}
    \centering
    \includegraphics[width=0.5\textwidth]{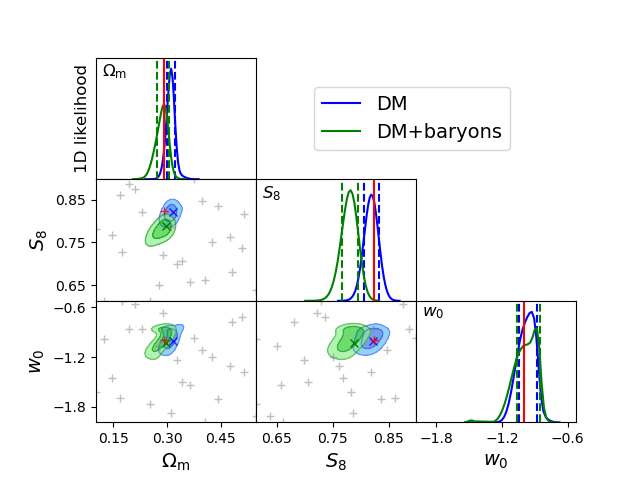}
    \caption{Same as Fig.~\ref{fig:forecasts_8_1} but for the combination of peak counts and $\gamma$-2PCF for a tomographic analysis with five redshift slices and including auto- and cross-terms between redshift slices.}
    \label{fig:forecasts_811_25}
\end{figure}

\new{Figures~\ref{fig:forecasts_8_1} to~\ref{fig:forecasts_811_25} show the 1 and $2\sigma$ contours of the marginalized 2D and 1D likelihoods for $S_8$, $\Omega_{\rm m}$, and $w_0$ for various configurations of the peak statistics: without tomography, with tomography, and combined with the $\gamma$-2PCF with tomography. In all these figures, the blue contours and curves correspond to the forecasts for the DM-only observations, while the data yielding the green contours are infused with the baryon bias. As already noted in \citet{Martinet+20}, we accurately recover the input parameter values for the DM-only case, with only a $\sim1\sigma$ bias on $\Omega_{\rm m}$ in some cases, which is due to the sampling of the initial parameter space (see the reference above for more details).}

\new{We first consider the nontomographic peak results in Fig.~\ref{fig:forecasts_8_1}.} We see a shift towards smaller values of $S_8$ ($\Delta S_8 = S_{8{\rm ,best}}^{\rm DM+Baryons} - S_{8{\rm ,best}}^{\rm DM}=-0.028 \,(-3.4$\%)) and $\Omega_{\rm m}$ ($\Delta \Omega_{\rm m} = -0.018 \,(-6.1$\%)) when including baryons. As expected, including baryonic feedback mimics the effect of having lower matter density/structure growth, the two being highly degenerate for cosmic shear. These results are fully consistent with the $\Omega_{\rm m}$ shift recently found by \citet{Coulton+20} for peaks measured in the BAHAMAS simulations, but are larger than those found in the earlier work of \citet{Osato+15}, who reported a $-1.5$\% shift in $\Omega_{\rm m}$ from simulations with weaker AGN feedback.  Finally, we see negligible changes in $w_0$, in part because the sensitivity to that parameter is quite low in the nontomographic case.

When including tomography (see Fig.~\ref{fig:forecasts_8_25}), the constraining power increases significantly. We measure a very similar effect to that seen in the nontomographic case for $S_8$ and $\Omega_{\rm m}$ but with slightly smaller shifts ($\Delta S_8 = -0.024 \,(-2.9$\%) and $\Delta \Omega_{\rm m} = -0.005 \,(-1.7$\%)). This behavior is also found in the DES-Y1 \magneticum\ mocks \citep[see][and Appendix~\ref{sec:app}]{Harnois-Deraps+20}: with tomography, the noise in each mass map increases and tends to wash out the effect of baryons. We also find a small positive shift in the maximum of the 1D likelihood for $w_0$ ($\Delta w_0 = 0.035 \,(3.5$\%)). However, we note  that this result is only supported by a small distortion of the likelihood, which otherwise agrees fairly well with the DM-only case. 
Although this effect could be physically motivated, it is likely due to degeneracies in the parameter space notably between $S_8$ and $w_0$, and it will be interesting to see if it persists in future analyses.

\begin{table*} 
\caption[]{Forecasts on the biases due to baryons in 100 deg$^2$ \textit{Euclid}-like mocks. The bias is defined as $\Delta S_8 = S_{8{\rm ,best}}^{\rm DM+Baryons} - S_{8{\rm ,best}}^{\rm DM}$, where the best estimates correspond to the maxima of the marginalized 1D likelihoods. Numbers in parenthesis show the results in percentage of the input value.} 
\centering 
\begin{tabular}{lccc} 
\hline 
\hline 
 & $\Delta S_8$ & $\Delta w_0$ & $\Delta \Omega_{\rm m}$ \\ 
\hline 
voids, no tomo. & $-0.029$ ($-3.5$\%) & $0.0$ ($0.0$\%) & $0.0$ ($0.0$\%)             \\ 
peaks, no tomo. & $-0.028$ ($-3.4$\%) & $0.0$ ($0.0$\%) & $-0.018$ ($-6.1$\%)         \\ 
1D $M_{\rm ap}$, no tomo. & $-0.023$ ($-2.8$\%) & $0.0$ ($0.0$\%) & $-0.006$ ($-2.2$\%)         \\ 
$\gamma$-2PCF, no tomo. & $-0.011$ ($-1.3$\%) & $0.0$ ($0.0$\%) & $-0.057$ ($-19.7$\%)       \\
\hline
voids, incl. tomo. & $-0.021$ ($-2.6$\%) & $0.0$ ($0.0$\%) & $0.003$ ($0.9$\%)          \\ 
peaks, incl. tomo. & $-0.024$ ($-2.9$\%) & $0.035$ ($3.5$\%) & $-0.005$ ($-1.7$\%)      \\ 
1D $M_{\rm ap}$, incl. tomo. & $0.007$ ($0.8$\%) & $0.048$ ($4.8$\%) & $-0.004$ ($-1.5$\%)        \\ 
$\gamma$-2PCF, incl. tomo. & $-0.034$ ($-4.2$\%) & $0.037$ ($3.7$\%) & $-0.035$ ($-11.9$\%)     \\
\hline
voids + $\gamma$-2PCF, incl. tomo. & $-0.042$ ($-5.1$\%) & $-0.051$ ($-5.1$\%) & $-0.014$ ($-4.7$\%)    \\ 
peaks + $\gamma$-2PCF, incl. tomo. & $-0.037$ ($-4.4$\%) & $0.047$ ($4.7$\%) & $-0.019$ ($-6.5$\%)      \\ 
1D $M_{\rm ap}$ + $\gamma$-2PCF, incl. tomo. & $-0.039$ ($-4.8$\%) & $0.049$ ($4.9$\%) & $-0.003$ ($-1.0$\%)      \\ 
\hline 
\end{tabular} 
    \label{tab:res}
\end{table*} 

If we now combine peaks with the $\gamma$-2PCF in the tomographic setup (Fig.~\ref{fig:forecasts_811_25}), the results are similar to those from peaks alone, but accentuated in amplitude. We find $\Delta S_8 = -0.037 \,(-4.4$\%), $\Delta \Omega_{\rm m} = -0.019 \,(-6.5$\%), and $\Delta w_0 = 0.047 \,(4.7$\%). Both peaks and $\gamma$-2PCF are affected in a similar manner by baryons and thus their combination shows a larger effect. Although one could have hoped that the impact of baryons would diminish when adding the information of the $\gamma$-2PCF that partly comes from larger scales, this study demonstrates on contrary that baryonic physics do not vanish in this combination and need to be accounted for in future surveys.

Table~\ref{tab:res} shows the forecast biases for all the different tested DVs and their combination with the $\gamma$-2PCF. Overall, baryons impact the different $M_{\rm ap}$ estimators in a similar manner, as expected from the similarities in how they affect each DV in Fig.~\ref{fig:DV}. We report smaller biases for voids than for peaks, as already noted in \citet{Coulton+20} in the nontomographical case, and confirm this trend when including tomography, with possibly a very low bias on $w_0$. When combined with the $\gamma$-2PCF we also note a difference of sign in the $w_0$ bias which could highlight a different sensitivity of voids to this parameter, but is more likely due to a residual small peak in the likelihood from the interpolation of the DV in this case. The lensing PDF presents a comparable bias to peaks with slightly lower shifts as well. Although voids are the least affected by baryons, this analysis shows that all estimators present biases of a few percent on at least one of the probed cosmological parameters. Considering the degeneracies between cosmological parameters, this highlights the necessity for accounting for baryons when modeling the dependence of nonGaussian $M_{\rm ap}$ estimators on cosmology. Because these results depend on the particular implementation of baryonic feedback processes in the \magneticum\ simulation, we run an additional test where we only infuse half of the DV baryonic bias to compute the cosmological forecasts. Although we cannot accurately model the response of the $M_{\rm ap}$ statistics to AGN feedback with only one simulation, this case likely corresponds to a much lower feedback amplitude. We find a bias on cosmological parameters which approaches the percent value in the tomographic case. This strong dependency of the cosmological parameters on the amplitude of the infused baryonic effect suggests that the impact of baryons could be mitigated by integrating a modeling of AGN feedback in the likelihood, a possibility which is not studied here.

Although these biases of  a few percent are worrisome for future Stage IV cosmic shear surveys, we note that they remain fairly small compared to the statistical precision of current surveys. Our $100$ deg$^2$ mocks at \textit{Euclid} depth include about ten million galaxies, a similar number to Stage III  surveys. In this case, the biases are always below $1\sigma$ for every parameter and configuration, except for the $S_8$ parameter when combining $M_{\rm ap}$ estimators and $\gamma$-2PCF where the bias can reach up to $2\sigma$. In Appendix~\ref{sec:app}, we tailor our mocks to the KiDS-450 and DES-Y1 surveys to verify the impact of baryons on the peak statistics analyses conducted in \citet{Martinet+18} and \citet{Harnois-Deraps+20}, respectively. Our findings validate the choice of neglecting baryons in current Stage III peak count analyses.

\section{Mitigating baryons with small scale cut}
\label{sec:cor}

In the case of $\gamma$-2PCF, the impact of baryons can be mitigated by discarding small scales which are the most affected, as seen from Fig.~\ref{fig:DV2pcf}. Although this process decreases the statistical precision as it removes part of the signal, it provides a gain in accuracy without needing to run computationally expensive hydrodynamical simulations. This trade-off was notably chosen by the DES collaboration in the first year data release \citep{Troxel+18}. \new{Recently, \citet{Taylor+18,Taylor+20} proposed an efficient way to cut the small-scale contribution to cosmic shear two-point estimators based on the nulling scheme presented in \citet{Bernardeau+14}.} Alternatively, baryons can be modeled with halo-based codes \citep[HMCode,][]{HMCode2020} or from libraries of power spectra \citep{vanDalen2020}, allowing the inclusion of smaller angular scales and increasing the statistical power as in \citet{Hildebrandt+17} and \citet{Huang2020}.

\begin{figure}
    \centering
    \includegraphics[width=0.5\textwidth]{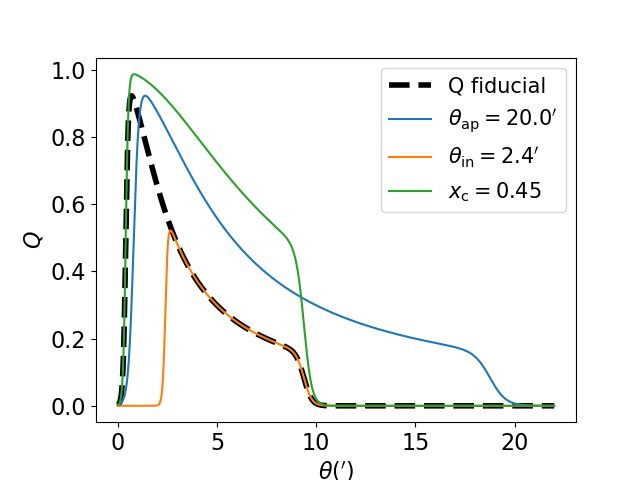}
    \includegraphics[width=0.5\textwidth]{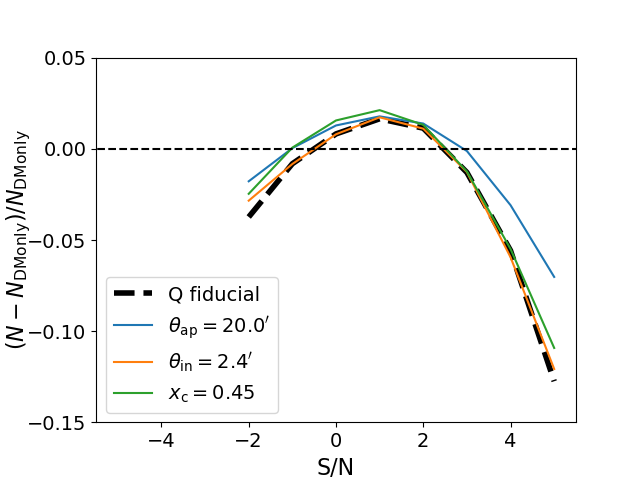}
    \caption{\textit{Top:} Profiles of the $Q$ filters defined via Eq.~(\ref{eq:Q}) and used to mitigate the impact of baryons by reducing the weight of small scales. \textit{Bottom:} Relative change in the $M_{\rm ap}$ peak counts due to baryons for the various mitigation setups.}
    \label{fig:miti}
\end{figure}

As we do not yet have access to a library of hydrodynamical simulations to estimate the baryonic bias, we follow the work of \citet{Weiss+19} and explore different possibilities to mitigate the effect of baryons on $M_{\rm ap}$ statistics. As suggested by their analysis, we vary the size of the aperture $\theta_{\rm ap}$ but we also investigate possible variations to the \citet{Schirmer+07} $Q$ filter shape entering Eq.~(\ref{eq:map}). In particular, we vary the aperture filter size $\theta_{\rm ap}$, the tilt parameter $x_{\rm c}$ , and the inner filter radius parameter $\theta_{\rm in}$.

The top part of Fig.~\ref{fig:miti} shows the impact of varying $\theta_{\rm ap}$, $\theta_{\rm in}$, and $x_{\rm c}$ on the shape of the $Q$ filter. The fiducial values that we use in the rest of the paper are $\theta_{\rm ap}=10'$, $\theta_{\rm in}=0.4'$, and $x_{\rm c}=0.15$. In this figure, we only present results when varying one parameter at a time for $\theta_{\rm ap}=20'$, $\theta_{\rm in}=2.4'$, and $x_{\rm c}=0.45$ to better highlight the effect of each parameter. However, we investigate multiple values in the ranges $3.3' \leq \theta_{\rm ap} \leq 106.7'$, $ 0.4' \leq \theta_{\rm in} \leq 3'$, and $0.05 \leq x_{\rm c} \leq 1$ and explore variations of all three parameters together. Increasing any of these three parameter values reduces the relative importance of the small scales in different manners. Increasing $\theta_{\rm ap}$ adds galaxies further away from the aperture center; increasing $\theta_{\rm in}$ removes galaxies close to the aperture center without distorting the general weighting; finally, increasing $x_{\rm c}$ up-weights distant galaxies without modifying the number of galaxies captured by the aperture. We re-compute the baryonic bias from $M_{\rm ap}$ constructed with the new $Q$ filters and observe, in the bottom part of Fig.~\ref{fig:miti}, that none of these methods is completely efficient at removing the impact of baryons. \new{This result is somewhat contradictory to our measurement that the effect of baryons in our simulations concentrates on the central few arcminutes of the $M_{\rm ap}$ profile around massive halos. This is because with mass map} statistics we cannot discard close galaxies as is done for $\gamma$-2PCF: we only reduce their weight when centered on them which does not prevent them from entering other apertures on a $\theta_{\rm ap}$ scale. We note a mild improvement when increasing $\theta_{\rm in}$ and $x_{\rm c}$. Increasing $\theta_{\rm ap}$ to $20.0'$ still decreases the impact of baryons by $\sim50$\% at the extreme S/N values, and only the largest $\theta_{\rm ap}>100'$ can bring it to zero, but the cost in precision is high, as we show next.

We apply these filter modifications to the measurements from the cosmology and covariance mocks, and carry out a full cosmological forecast  for each of these. In
Table~\ref{tab:rescut}, we present the variations of the bias on the inferred cosmology for these different configurations relative to the fiducial $Q$ filter. We also report the change in the forecast precision due to the reduction in small-scale information. We show these results for peaks in a configuration without tomography but we find similar behavior for other $M_{\rm ap}$ estimators and including tomography: the bias due to baryons is reduced (e.g., $\Delta S_8^{} / \Delta S_8^{Q\,{\rm fid.}} < 1.0$), but at the cost of a reduction in the statistical precision ($\delta S_8^{} / \delta S_8^{Q\,{\rm fid.}} > 1.0$). Quantitatively, we find with $\theta_{\rm ap}=20'$ that the bias is reduced by a factor of almost two and three on $S_8$ and $\Omega_{\rm m}$ respectively, but with a loss of respectively $12$\% and $17$\% on the forecast precision, and a loss of $15$\% on $w_0$. The two other variations are less efficient but also retain more of the cosmological information. The cut at $\theta_{\rm in}=2.4'$ decreases the bias by $\sim30$\% at the cost of a $5$\%, $\sim10$\%, and $\sim20$\% wider statistical precision on $S_8$, $w_0$, and $\Omega_{\rm m}$, respectively. We find similar results for all the $Q$ filter configurations using higher parameter values than the fiducial: the gain in accuracy is always balanced by a significant loss in precision. However, when using smaller values of $\theta_{\rm ap}$, $\theta_{\rm in}$, or $x_{\rm c}$, the impact of baryons is increased because of the larger contribution of the small-scale baryonic features but the constraining power is also degraded as we chose the fiducial $Q$ filter to maximize the forecast precision in \citet{Martinet+20}.

\begin{table*} 
\caption[]{Changes in the measured bias due to baryons and in the associated forecast precision for various mitigation schemes in the case of $M_{\rm ap}$ peaks without tomography. The comparison is performed with respect to the fiducial $Q$ filter used in the rest of the paper with $\theta_{\rm ap} = 10'$, $\theta_{\rm in} = 0.4'$, and $x_{\rm c} = 0.15$. We vary one parameter at a time, the two others being held to their fiducial values. ``$\Delta$'' refers to the bias due to baryons, and ``$\delta$'' to the $1\sigma$ precision forecast.} 
\centering 
\begin{tabular}{lcccccc} 
\hline 
\hline 
 & $\Delta S_8^{} / \Delta S_8^{Q\,{\rm fid.}}$ & $\Delta w_0^{} / \Delta w_0^{Q\,{\rm fid.}}$ & $\Delta \Omega_{\rm m}^{} / \Delta \Omega_{\rm m}^{Q\,{\rm fid.}}$ & $\delta S_8^{} / \delta S_8^{Q\,{\rm fid.}}$ & $\delta w_0^{} / \delta w_0^{Q\,{\rm fid.}}$ & $\delta \Omega_{\rm m}^{} / \delta \Omega_{\rm m}^{Q\,{\rm fid.}}$ \\ 
\hline 
peaks, no tomo.          & & & & & &  \\ 
$\theta_{\rm ap} = 20'$  & $0.47$ & $1$ & $0.39$ & $1.12$ & $1.15$ & $1.17$ \\ 
$\theta_{\rm in} = 2.4'$ & $0.65$ & $1$ & $0.64$ & $1.05$ & $1.09$ & $1.19$ \\ 
$x_{\rm c} = 0.45$       & $0.76$ & $1$ & $0.59$ & $1.05$ & $1.11$ & $1.15$ \\ 
\hline 
\end{tabular} 
    \label{tab:rescut}
\end{table*} 

Overall, none of the small-scale cuts we applied to mitigate baryonic effects are able to decrease the bias below a few percent accuracy whilst preserving strong statistical precision. Using the effective scale of $\theta_{\rm ap}x_{\rm c}=16'$ recommended in the analysis of \citet{Weiss+19} for \textit{Euclid}-like mocks, we confirm that the bias becomes consistent with zero. However, such large smoothing scale results in a decrease in the constraining power by a factor of more than $2$, $1.5$, and $3$ on $S_8$, $w_0$, and $\Omega_{\rm m}$, respectively, motivating a full forward-model approach of the baryonic bias.

\section{Conclusion}
\label{sec:ccl}

In this paper we investigate the impact of baryons on various $M_{\rm ap}$ statistics: peaks, voids, and the lensing PDF. Baryonic physics is modeled with the state-of-the-art \magneticum\  hydrodynamical simulations, and its impact on the data vector is propagated into full  cosmological forecasts on $S_8$, $w_0$, and $\Omega_{\rm m}$, for a Stage-IV lensing survey. The likelihood sampling exploits the cosmological pipeline of \citet{Martinet+20}, which is based on the SLICS and cosmo-SLICS DM-only simulations. Our results are summarized below:

\begin{itemize}

    \item[$\bullet$] Baryons bias the measured $M_{\rm ap}$ estimators by about $5-10$\% on most S/N bins, notably decreasing the number counts at extreme S/N values, while increasing the number of intermediate S/N  features. This is a direct consequence of strong baryonic feedback, which dilutes the density profile of massive halos and decreases their S/N in the $M_{\rm ap}$ map.\\
    
    \item[$\bullet$] In our Stage-IV survey setup without tomography, the baryonic feedback propagates into a negative bias of about $-3$\% on $S_8$ for every estimator. The bias on $\Omega_{\rm m}$ depends on the estimator and ranges from zero for voids to $-6$\% for peaks. When including a tomographic decomposition with five redshift slices including cross-tomographic bins, these biases are slightly lowered, likely because of the increased shape noise in each tomographic slice, but remain of the order of a few percent. We observe positive bias in our tomographic setup of the order of $3-5$\% on $w_0$, although it is not clear at the moment whether this has a physical origin or is caused by parameter degeneracies that are not fully captured by our likelihood.\\
    
    \item[$\bullet$] Biases on all parameters are increased to $\sim \!\! 5$\%  when combining any $M_{\rm ap}$ statistics with $\gamma$-2PCF in the tomographic analysis. This combined analysis is maximally affected as the contributions of baryons are in the same direction and of similar amplitude for the individual probes.\\
    
    \item[$\bullet$] After investigating a range of scale cuts on the $M_{\rm ap}$ statistics, we find that it is difficult to efficiently lower the impact of baryons without significantly degrading the statistical power. 
    In line with \citet{Weiss+19}, we find that only an overly large aperture size could lower the bias to sub-percent level. A large portion of the signal is lost with such filtering, leading to less competitive cosmological constraints.\\
    
\end{itemize}  

We built \magneticum\ mocks to measure the impact of baryons on peak statistics analyses of current stage III surveys, namely in KiDS-450 \citep{Martinet+18} and DES-Y1 \citep{Harnois-Deraps+20}, and show that it remains below the statistical uncertainties associated to these surveys. However, this will not be the case for future Stage IV surveys for which baryons need to be accounted for in order to reach percentage-level precision whilst remaining unbiased.

In this article, we present a correction scheme based on the DV: we measure a corrective factor from hydrodynamical simulations and apply it to mock observations in order to estimate how biased would become cosmological constraints in cases where this step was omitted. However, we note that these results fully depend on the amplitude of the baryonic feedback modeled by the \magneticum\ simulations, which is still uncertain as seen from the scatter between the different state-of-the-art simulations. A more accurate correction would consist in modeling the impact of baryons on $M_{\rm ap}$ statistics from a set of hydrodynamical simulations with various feedback amplitude values, and to marginalize over the extra free parameters when computing the cosmological constraints. \citet{Coulton+20} recently showed the feasibility of this approach using the BAHAMAS simulations run with three different amplitudes of the AGN feedback. The baryonification method described in \citet{Schneider+15a} is particularly suited to such an analysis and is therefore a promising tool to design future sets of $N$-body simulations that explore both the cosmology dependence and the response to baryons of nonGaussian statistics.

\begin{acknowledgements}
We thank our KiDS and \textit{Euclid} collaborators for useful discussions. NM acknowledges support from a fellowship of the Centre National d'Etudes Spatiales (CNES). TC is supported by the INFN INDARK PD51 grant and by the PRIN-MIUR 2015W7KAWC grant. JHD acknowledges support from an STFC Ernest Rutherford Fellowship (project reference ST/S004858/1). CG acknowledges support from PRIN MIUR 2017 WSCC32 ``Zooming into dark matter and proto-galaxies with massive lensing clusters.'' KD acknowledges support by the Deutsche Forschungsgemeinschaft (DFG, German Research  Foundation)  under  Germany's  Excellence  Strategy  -- EXC-2094 -- 390783311. \magneticum\ has been run using the ``Leibniz-Rechenzentrum'' with CPU time assigned to the Project “pr86re” and “pr83li”. The SLICS numerical simulations can be found at \url{http://slics.roe.ac.uk/}, while the cosmo-SLICS can be made available upon request. Computations for these $N$-body simulations were enabled by Compute Ontario (\url{www.computeontario.ca}), Westgrid (\url{www.westgrid.ca}) and Compute Canada (\url{www.computecanada.ca}). This work has been carried out thanks to the support of the OCEVU Labex (ANR-11-LABX-0060) and of the Excellence Initiative of Aix-Marseille University - A*MIDEX, part of the French ``Investissements d'Avenir'' program.\\
~\\
All authors contributed to the development of this paper. NM (lead) conducted the analysis. TC and JHD (both co-leads) equally participated by producing the \magneticum\ and the SLICS and cosmo-SLICS mocks respectively. EJ contributed to the analysis pipeline while CG and KD prepared the initial configuration of the \magneticum\ mocks.
\end{acknowledgements}

\bibliographystyle{aa}
\bibliography{spe2}

\appendix

\section{Correcting for baryons in KiDS and DES}
\label{sec:app}

In this section we build \magneticum\ mocks for current Stage III surveys to investigate the effect of baryons on peak counts in recent analyses. We first revisit the results of \citet{Martinet+18} on the KiDS-450 data, and then present a comparison with those obtained  for the DES-Y1 analysis of \citet{Harnois-Deraps+20}.

\subsection{Estimation of the bias for KiDS-450 \citep{Martinet+18}}

We follow the same procedure as in \citet{Martinet+18} to create KiDS mocks, but now we additionally include baryonic physics and examine the impact on the inferred cosmology. In short, we use the KiDS-450 redshift distribution calibrated with the direct spectroscopic method \citep[DIR;][]{Hildebrandt+17} for a single tomographic slice including all galaxies with photometric redshifts $0.1<z<0.9$ (no tomography is applied). The galaxy density of the mocks is approximately $7.5$ arcmin$^{-2}$ and we tile the full $450$ deg$^2$ by repeating our $100$ deg$^2$ simulation mocks. In this process we set the positions and the amplitude of the intrinsic ellipticities of the mock galaxies to that of the observed data. We use independent simulated shear values from the ten lines of sight reconstructed in the \magneticum\ and apply the same five different random rotations of the intrinsic ellipticities as in the KiDS mocks used to compute the model. We build $M_{\rm ap}$ maps with the same \citet{Schirmer+07} filter with an aperture size $\theta_{\rm ap}=12.5'$. The effect of baryons is measured as the ratio between the mean distribution of peaks in the $M_{\rm ap}$ maps built from the $50$ hydrodynamical and DM-only mocks, in $12$ bins of S/N ranging from $0$ to $4$. It is then infused as a multiplicative factor applied to the observed data. In this approach we aim to remove the effect of baryons from the data rather than modifying the model (both methods are equivalent). The correction is then propagated to the cosmological constraints using the same pipeline as in \citet{Martinet+18}, which we recall is built from the $157$ \citet{DH10} $N$-body simulations that pave the $\sigma_8 - \Omega_{\rm m}$ plane, plus 35 at the fiducial cosmology for the covariance matrix estimation. When including various light-cones and shape noise realizations, the model is constructed from a total of $3925$ mocks, with an additional $175$ {\it pseudo}-independent mocks for the covariance matrix.

\begin{figure}
    \centering
    \includegraphics[width=0.5\textwidth]{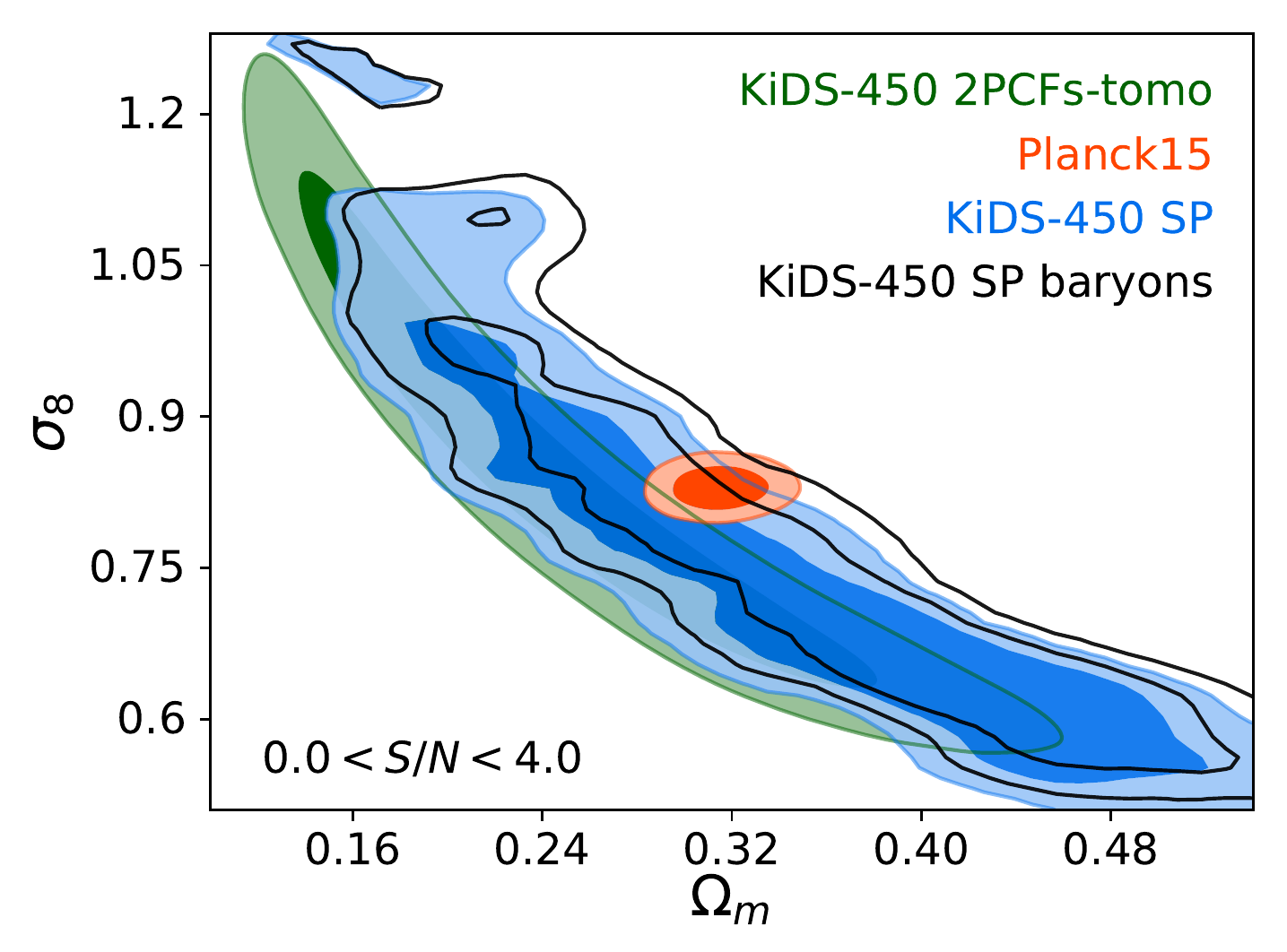}
    \caption{Effect of baryons on the KiDS-450 $M_{\rm ap}$ peaks cosmological constraints of \citet{Martinet+18}. Blue and black contours correspond to the $1$ and $2\sigma$ constraints without and with accounting for baryons respectively. Green and red contours represent the best KiDS-450 tomographic $\gamma$-2PCF constraints \citep{Hildebrandt+17} and Planck cosmic microwave background constraints \citep{PlanckXIII} that were available at the time.}
    \label{fig:kids}
\end{figure}

The effect of baryons on the KiDS-450 peak counts analysis is shown in Fig.~\ref{fig:kids}. The blue contours correspond to the $1$ and $2\sigma$ constraints presented in Figure 7 of \citet{Martinet+18} and which do not include systematic errors. The black lines show the constraints including the effect of baryons. We see a positive shift in both $\sigma_8$ and $\Omega_{\rm m}$ resulting in a change of the structure growth parameter $\Delta S_8 = 0.021~(2.8\%)$. We note that the bias is positive in this case: given a fixed observed DV which already includes baryons, including them in the model results in a higher $S_8$ cosmology, i.e. the number of large S/N peaks in the model is reduced because of the baryons. This is in contrast with the simulation-based approach in the rest of the article where we estimate the bias from infusing baryons to the DM-only observation. The effect in KiDS is lower than the $3.4$\% measured in the \textit{Euclid}-like mocks, likely because of the larger noise with the lower galaxy density of KiDS. It is nevertheless larger than the $2.3$\% correction used in \citet{Martinet+18} and derived from the simulations of \citet{Osato+15}, as expected from the weaker AGN feedback implemented in the latter study. When compared to the statistical precision of the KiDS-450 analysis, the $S_8$ shift due to baryons remains small with a value of $0.27\sigma$. With this updated baryon model, we revise the $S_8$ constraints from \citet{Martinet+18}, including the other sources of systematic error described therein: multiplicative shear bias, mean photometric redshift, intrinsic alignment, and shear-position coupling. As the impact of baryons is now fully modeled, and because the constraining power is identical with or without baryons, we no longer need to inflate the uncertainty on $S_8$. We find $S_8 = 0.788^{+0.057}_{-0.056}$, a result slightly closer to the Planck estimate \citep[][the red contours in Fig. \ref{fig:kids}]{PlanckXIII} than the $S_8 = 0.750^{+0.059}_{-0.058}$ previously reported.


\subsection{Estimation of the bias for DES-Y1 \citep{Harnois-Deraps+20}}

The \magneticum\ simulations have recently been used in \citet{Harnois-Deraps+20} to investigate the impact of baryons on the peak statistics analysis of the DES-Y1 data. These mocks use the DIR-calibrated redshift distributions used in the cosmic shear  re-analysis by \citet{Joudaki+20}, carried out  in four photometric redshift bins between $0.2$ and $1.3$. 
We review the measurement of the baryons bias here and compare the results to those  obtained in the previous sections.

As described in \citet{Harnois-Deraps+20}, we tile the full DES-Y1 survey ($1321$ deg$^2$) with our $100$ deg$^2$ \magneticum\  mocks and fix the galaxy positions and intrinsic ellipticity amplitudes to that of the data. We create $100$ mock surveys for the DM-only and the hydrodynamical \magneticum\ simulations to lower the sample variance and shape noise. In this analysis, the model and the covariance matrix are evaluated from the same cosmo-SLICS and SLICS $N$-body simulations used in the main part of this article respectively, improving in accuracy from the  \citet{DH10} $N$-body simulations used in the KiDS-450 peak count analysis. 
The $M_{\rm ap}$ are computed with an aperture size of $\theta_{\rm ap}=12.5'$ in the four auto-tomographic bins and in the cross-bins. The DV is the concatenation of the peak distributions in $12$ bins between $0<{\rm S/N}<4$ for all tomographic configurations. A total of $2600$ mock surveys are used for the $w$CDM peak count model, and $1240$ for the covariance matrix.

\begin{figure}
    \centering
    \includegraphics[width=0.5\textwidth]{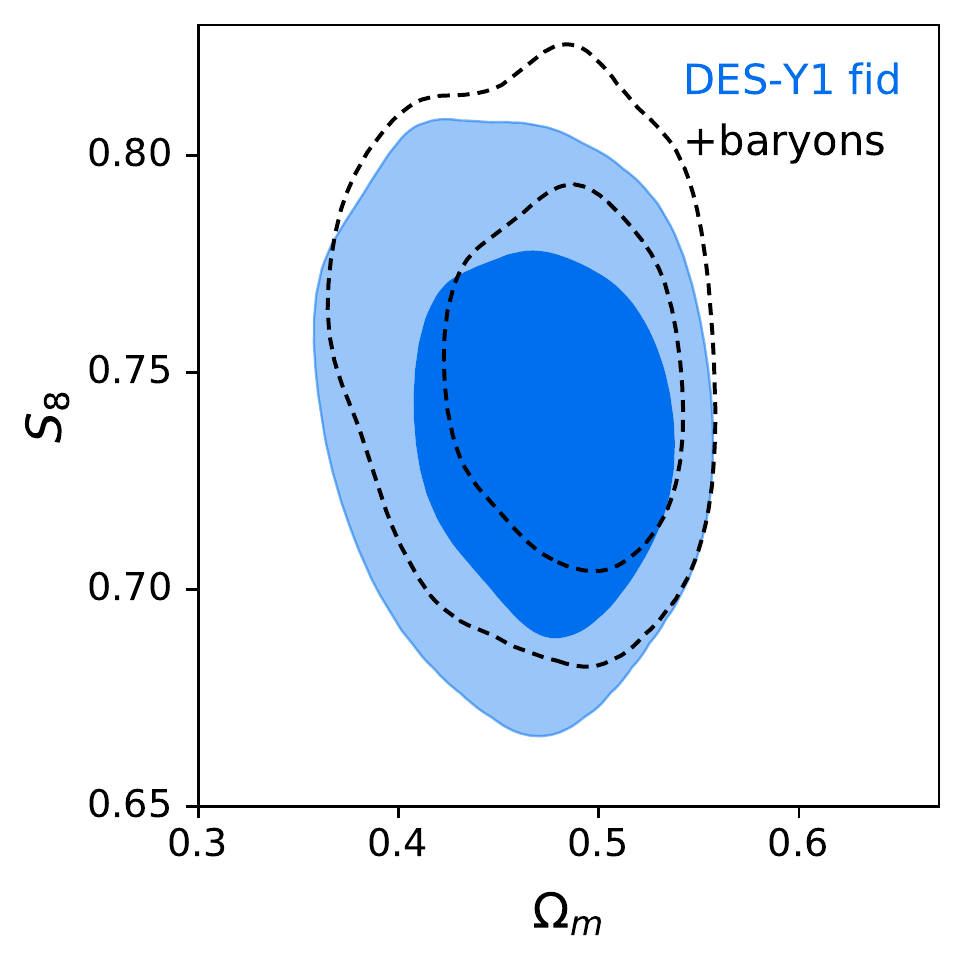}
    \caption{Effect of baryons on the DES-Y1 $M_{\rm ap}$ peaks cosmological constraints of \citet{Harnois-Deraps+20}. Blue and black contours correspond to the $1$ and $2\sigma$ constraints without and with accounting for baryons respectively.}
    \label{fig:des}
\end{figure}

The impact of baryons on the tomographic constraints from $M_{\rm ap}$ peak statistics in DES-Y1 are shown in Fig.~\ref{fig:des}. The blue contours correspond to the $1$ and $2\sigma$ constraints on the $S_8$ and $\Omega_{\rm m}$ parameters after marginalisation over the photometric redshift and the shear calibration uncertainties, while the black contours further include the correction due to baryonic physics. As expected, we find again a positive shift towards larger values of $S_8$ and $\Omega_{\rm m}$. Quantitatively, the shift is of $\Delta S_8 = 0.013~(1.8\%)$, lower than the $2.9$\% found with the \textit{Euclid}-like mocks, again due to the lower galaxy density in the DES-Y1 data, which is $\sim 5$ arcmin$^{-2}$. In terms of statistical precision, this bias corresponds to a $0.32\sigma$ shift and can be safely ignored in the DES-Y1 analysis. \citet{Harnois-Deraps+20} explored other sources of systematics in their analysis, and show in their Figure 17 that baryons and intrinsic alignments are the two most important effects, while source-lens clustering and uncertainty in the nonlinear growth of structure could be safely ignored.

\end{document}